\documentclass[sigconf]{acmart}

\AtBeginDocument{%
  }

\usepackage{multicol}
\usepackage{microtype}
\usepackage{graphicx}
\usepackage{booktabs}
\usepackage{tabularx}
\usepackage{longtable}
\usepackage{xspace}
\usepackage[capitalise]{cleveref}
\usepackage{balance}
\usepackage[colorinlistoftodos,prependcaption,textsize=tiny]{todonotes}

\newcommand{\toolname}{\textsc{\mbox{IntelliGame}}\xspace}
\newcommand{\control}{\emph{control}\xspace}

\newcommand{\treatment}{\emph{treatment}\xspace}

\newcommand{\universityname}{Politecnico di Torino}

\copyrightyear{2024}
\acmYear{2024}
\setcopyright{acmlicensed}\acmConference[IDE '24]{2024 First IDE Workshop}{April 20, 2024}{Lisbon, Portugal}
\acmBooktitle{2024 First IDE Workshop (IDE '24), April 20, 2024, Lisbon, Portugal}
\acmDOI{10.1145/3643796.3648466}
\acmISBN{979-8-4007-0580-9/24/04}


\begin{document}

%%
%% The "title" command has an optional parameter,
%% allowing the author to define a "short title" to be used in page headers.
%\title{Empirical experiment in Javascript in IntelliGame}
%\title{Powering Up JavaScript Unit Testing with IntelliGame: An Experience Report}
%Coding Challenges Redefined: IntelliGame in JavaScript Unit Testing
\title{IntelliGame in Action: An Experience Report on \\Gamifying JavaScript Unit Tests}

%%
%% The "author" command and its associated commands are used to define
%% the authors and their affiliations.
%% Of note is the shared affiliation of the first two authors, and the
%% "authornote" and "authornotemark" commands
%% used to denote shared contribution to the research.
\author{Philipp Straubinger}
\authornote{Both authors contributed equally to this research.}
\affiliation{%
	\institution{University of Passau}
	\city{Passau}
	\country{Germany}
}

\author{Tommaso Fulcini}
\authornotemark[1]
\affiliation{%
	\institution{Politecnico di Torino}
	\city{Torino}
	\country{Italy}
}

\author{Gordon Fraser}
\affiliation{%
	\institution{University of Passau}
	\city{Passau}
	\country{Germany}
}

\author{Marco Torchiano}
\affiliation{%
	\institution{Politecnico di Torino}
	\city{Torino}
	\country{Italy}
}


\begin{abstract}

%This paper explores the integration and assessment of \toolname, a gamification plugin originally crafted for Java development, within the domain of JavaScript unit testing. The objective is to affirm IntelliGame's efficacy in the context of JavaScript and offer insights into the experiment's design. 
This paper investigates the integration and assessment of \toolname, a gamification plugin initially designed for Java development, within the realm of JavaScript unit testing. We aim to verify the generalizability of \toolname to JavaScript development and to provide valuable insights into the experiment's design. For this, we first customize \toolname for JavaScript, and then conduct a controlled experiment involving 152 participants utilizing the Jest testing framework, and finally examine its influence on testing behavior and the overall developer experience. The findings from this study provide valuable insights for improving JavaScript testing methodologies through the incorporation of gamification.

\end{abstract}

\maketitle

\section{Introduction}

%In our previous work~\cite{DBLP:conf/icse/Straubinger024}, we introduced \toolname, a plugin for the popular IntelliJ IDEA. The plugin introduces 27 different achievements, each with incremental levels, providing positive feedback when developers exhibit commendable testing behavior. The achievements cover various aspects of software testing, such as test execution, coverage evaluation, debugging, and test refactoring. 
% MTk: rearranged the above as follows:
\toolname~\cite{DBLP:conf/icse/Straubinger024} is a plugin for the popular IntelliJ IDEA, which enables gamification covering various aspects of software testing, such as test execution, coverage evaluation, debugging, and test refactoring. It introduces 27 different achievements, each with incremental levels, providing positive feedback when developers exhibit commendable testing behavior.
\toolname provides real-time analysis of developer interactions, notifications, and a user interface within IntelliJ to display achievements and progress.

In its original evaluation~\cite{DBLP:conf/icse/Straubinger024}, we conducted a controlled experiment with 49 participants to study the impact of \toolname on testing behavior. The experiment involved a Java programming task with participants divided into different groups,
%, with the participants divided into four groups: \control, \notifications, \treatment, and \maximizing. The experiment
aiming to assess the impact of the gamified environment on testing behavior, test suites, achievement levels, code functionality, and the overall developer experience.
The results showed clearly that \toolname had a positive effect on testing behavior.
%, with achievements playing a more significant role than notifications.
Participants using \toolname wrote more tests, achieved higher code coverage and mutation scores, ran tests more frequently, and implemented functionality earlier compared to the \control group. The impact on resulting test suites was substantial, and achievements correlated positively with various testing metrics.

The goal of this paper is to build upon these promising results achieved with \toolname for Java environments, and to assess whether these outcomes can be replicated on a larger scale and in a different context. In particular, the extension of the positive impact of achievements in the IDE to various programming languages remains unexplored. Thus we shifted the focus of this study to JavaScript (JS), recognized as the most used language as of 2023~\cite{statista}. With the growing importance of JS, there is an increasing demand to instruct and practice language-specific testing activities.
Expanding beyond the previous sample, this study engages a broader audience with multiple tasks, allowing for more extensive code development and testing. This expansion, coupled with the shift to a different programming language, considers both generalizability and effectiveness beyond a short-term context.

The objective of this paper is to provide an in-depth description of the experiment's design, including challenges encountered, lessons learned, and recommendations for future iterations of JS-based experiments. Notably, there is a gap in the existing literature regarding programming experiments centered on an open-source JavaScript target~\cite{sun2017analysis}, emphasizing a fully replicable and systematic approach. In  \cref{sec:background} we provide some background about the topic of the paper with also some examples of related works, in\cref{sec:implementation} the implementation approach for the JavaScript version of \toolname is described, while \cref{sec:experiment} shows the selection process for the project as a subject is meticulously detailed, shedding light on the rationale behind the choice.

\section{Background and Related Work} \label{sec:background}

Compared to writing code, testing is often viewed as  less rewarding by software developers, requiring significant effort without always resulting in due recognition from management~\cite{10301252}. To address issues related to motivation and acknowledgment of testers' contributions, various solutions have been proposed, encompassing both extrinsic and intrinsic motivators. Gamification, a rising trend in this domain, has experienced substantial growth over the past decade~\cite{MLR}, even leading to dedicated conferences in the academic sector.

Defined as \textit{the use of game elements in a non-playful context} by Deterding et al.~\cite{Deterding}, gamification involves creating a playful environment in which users can accomplish their daily tasks. Completing daily goals within such an environment allows users to benefit from the motivators associated with the designed game elements, making the underlying activity more engaging and satisfying.

Software testing has been successfully gamified in particular in the context of education. For example, Code Defenders~\cite{code_defenders} is a gamified application to teach mutation testing concepts in an academic setting, where students assume either of two roles: attackers or defenders. Users from both sides engage in a mutual challenge on a shared Java class, applying mutation testing. The attackers' goal is to create code mutants (variants of the class with the same functionality), while the defenders' objective is to enhance the existing test suite with new test cases to detect the code mutants. The gamified platform incorporates mechanics and dynamics that integrate social interaction, creativity, and competition.

Gamification has also been applied in a practical rather than educational context. For example, Coppola et al.~\cite{10375892} propose a framework to gamify exploratory GUI testing using a Capture and Replay tool for web applications. The gamification layer introduces common game elements such as scores and leaderboards, as well as novel elements not previously explored in the literature, including dynamic visual bug injection with related visual feedback and scores, along with a progress bar tracking the tester's exploration, i.e., the achieved widget coverage of a specific web page.

As indicated in a recent survey on gamified software testing~\cite{MLR}, current trends in the literature show that unit testing is the most targeted level, with the most focused testing phase being test creation and execution. Popular techniques considered by researchers and practitioners include Mutation, Black Box, and White Box testing. While \toolname initially emerged as a unit test tool, its achievement-based structure allows for extensibility across various testing dimensions, introducing achievements specific to each. %The following section will delve into the adaptation to JavaScript.

\section{Implementation} \label{sec:implementation}

\begin{figure}[t]
    \centering
    \includegraphics[width=\linewidth]{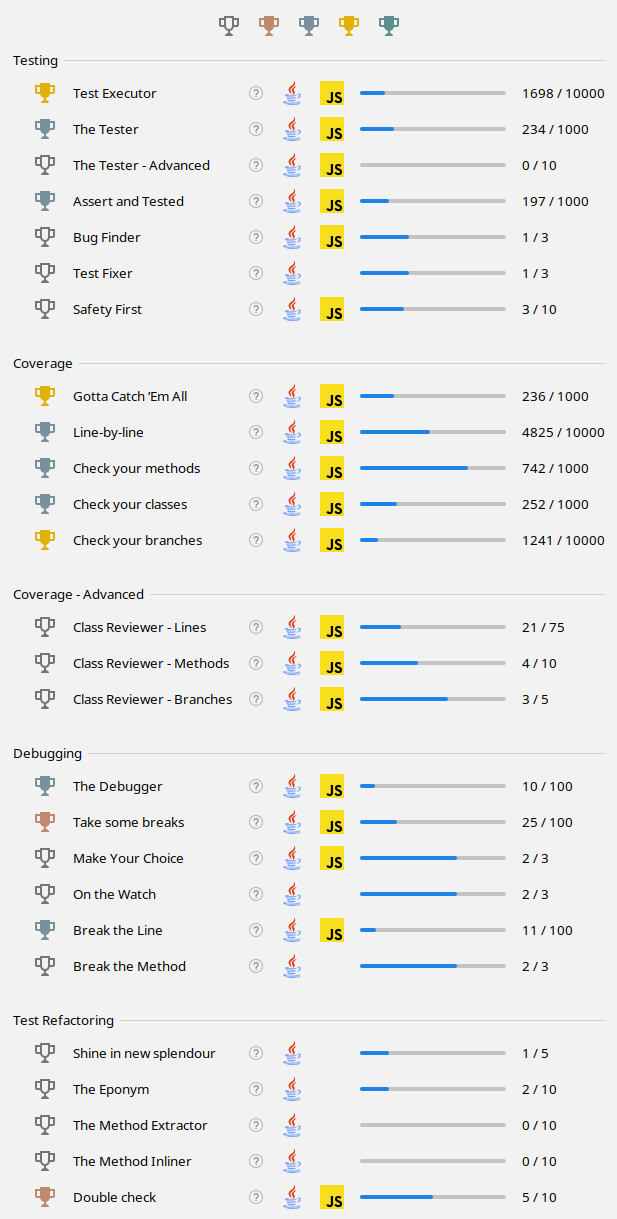}
    \caption{IntelliJ window showing achievements and progress}
    \label{fig:achievements}
\end{figure}

To enable JavaScript support in \toolname, we transitioned from using IntelliJ Community Edition to IntelliJ Ultimate\footnote{\url{https://www.jetbrains.com/products/compare/?product=idea&product=idea-ce}}. This shift was necessary because JavaScript support is exclusively available in the latter. While we managed to adapt most of the existing test achievements for JavaScript, some proved challenging to port, and a few were even deemed impossible.

Adjusting for test execution tracking, such as the number of tests and assertions, was straightforward, thanks to the Jest testing framework's syntax\footnote{\url{https://jestjs.io/}}. However, obtaining coverage information posed a greater challenge. The built-in test engine of IntelliJ lacks the same coverage details for JavaScript as it does for Java, such as covered branches or methods.

To address this, we devised a specialized test execution configuration that utilizes Jest to output overall coverage information to the console. We extract and employ this information for our achievements. Additionally, Jest writes coverage details to a JSON file, offering comprehensive class-specific information that we leverage for the remaining achievements. In summary, we extract JavaScript coverage information from three sources: the built-in test engine, the console, and summary files.

Another obstacle arose with the JavaScript debugger, which lacks support for method breakpoints and field watchpoints, required by the implementation of some of the achievements. Unfortunately, we could not find suitable alternatives for these features. Similarly, most of the refactoring-related achievements, implemented using RefactoringMiner for Java\footnote{\url{https://github.com/tsantalis/RefactoringMiner}}, lacked equivalent tools for JavaScript.

Despite facing challenges, we successfully transferred 19 achievements out of 26 from Java to JavaScript. These achievements can be identified by a small JavaScript symbol (see \cref{fig:achievements}). Each achievement is represented by a trophy indicating the level, a progress bar showing the current progress towards the next level, and a description detailing the progress required for reaching the next level, accessible by hovering over the question mark. For further details and explanations of the achievements, refer to related work~\cite{DBLP:conf/icse/Straubinger024}.

\section{Experiment} \label{sec:experiment}

The primary aim of the experiment was to replicate the conditions outlined in the original validation paper \cite{DBLP:conf/icse/Straubinger024}, adhering to the guidelines set by Jedlitschka et al.~\cite{jedlitschka2005reporting}. However, as the goal was not to assess the effectiveness of the tool but rather to test its applicability with a different programming language and subjects, we introduced some modifications to the experimental approach.

\subsection{Participant Selection}

We invited graduate students from the Software Engineering course at \universityname to participate voluntarily in the experiment. Successful participation earned students additional points for their Software Engineering exam, with the assurance that their performance would not impact their bonus points to avoid bias.
All participating students possessed prior JavaScript testing experience, as the Software Engineering course extensively covered unit testing using the Jest framework.

The sample of participants in this study comprised 152 individuals, with 85\% identifying as male and the remaining 15\% as female. Over 90\% fell within the age range of 22 to 25, being graduate students. %This demographic composition is unsurprising given that the experiment took place within a university course. 
The participants were randomly assigned to one of two groups: \treatment and  \control.

The majority of participants had less than three months of experience with JavaScript, and it is worth noting that the \treatment group had less experience compared to the \control group with 59\% and 47\% having less than three months of experience, respectively. Additionally, 96\% of participants in both groups had less than three months of experience with Jest. In fact, the majority of participants were introduced to JavaScript and Jest during the course's lectures.

\subsection{Project Selection}

Initially, with a shift in the reference programming language, the need arose to select an alternative project, departing from the \textit{FixedOrderComparator} Java class used in the original study~\cite{DBLP:conf/icse/Straubinger024}.

Our criteria for project selection involved finding a subject that presented both challenge and feasibility within a 150-minute timeframe. We sought a real existing project to capture the complexity of a real-world scenario, emphasizing documentation, a comprehensive test suite, and modularity. Modularity was crucial, allowing participants to choose among modules, and deciding whether to tackle more complex or simpler functions.

The selection process began by exploring JavaScripting\footnote{\url{https://www.javascripting.com/}}, a repository of publicly available JavaScript projects. Given the extensive collection, including popular frameworks like React, React Native, Vue, and Angular, we narrowed down our search to importable lightweight libraries with standardized functionalities. We focused on the \emph{miscellaneous} category for a subject meeting our criteria.

The first project suiting our needs was \emph{Date Fns}\footnote{\url{https://github.com/date-fns/date-fns}}, an open-source project featuring 244 functions related to the date data type. Functions ranged from trivial tasks like determining the order between two dates to more complex operations like formatting data with ISO or RFC representation. The project, widely used (over two million times) and well-maintained by 300+ contributors with over a thousand forks, met our requirements.

The project, however, was developed in TypeScript (TS) rather than JS. Notwithstanding the difference between the two programming languages, given the high project usage and reliability, the full interoperability of JS code in TS and the possibility of directly transposing TS code into JS, was deemed that this language difference was not a reason for exclusion. Moreover, we consider it valuable for the reader to discuss in section 5.2 the problems encountered and the solutions put in place to resolve them.  

For these reasons, we used the JS version obtained through transpilation~\cite{DBLP:journals/pacmpl/WangKNBS23}. The seamless interoperability between the two languages facilitated the conversion of functions from TS to JS, made possible with the help of the open-source automatic tool transform\footnote{\url{https://github.com/ritz078/transform}}.

\subsection{Experiment Task}

\begin{figure}[t]
	\includegraphics[width=\linewidth]{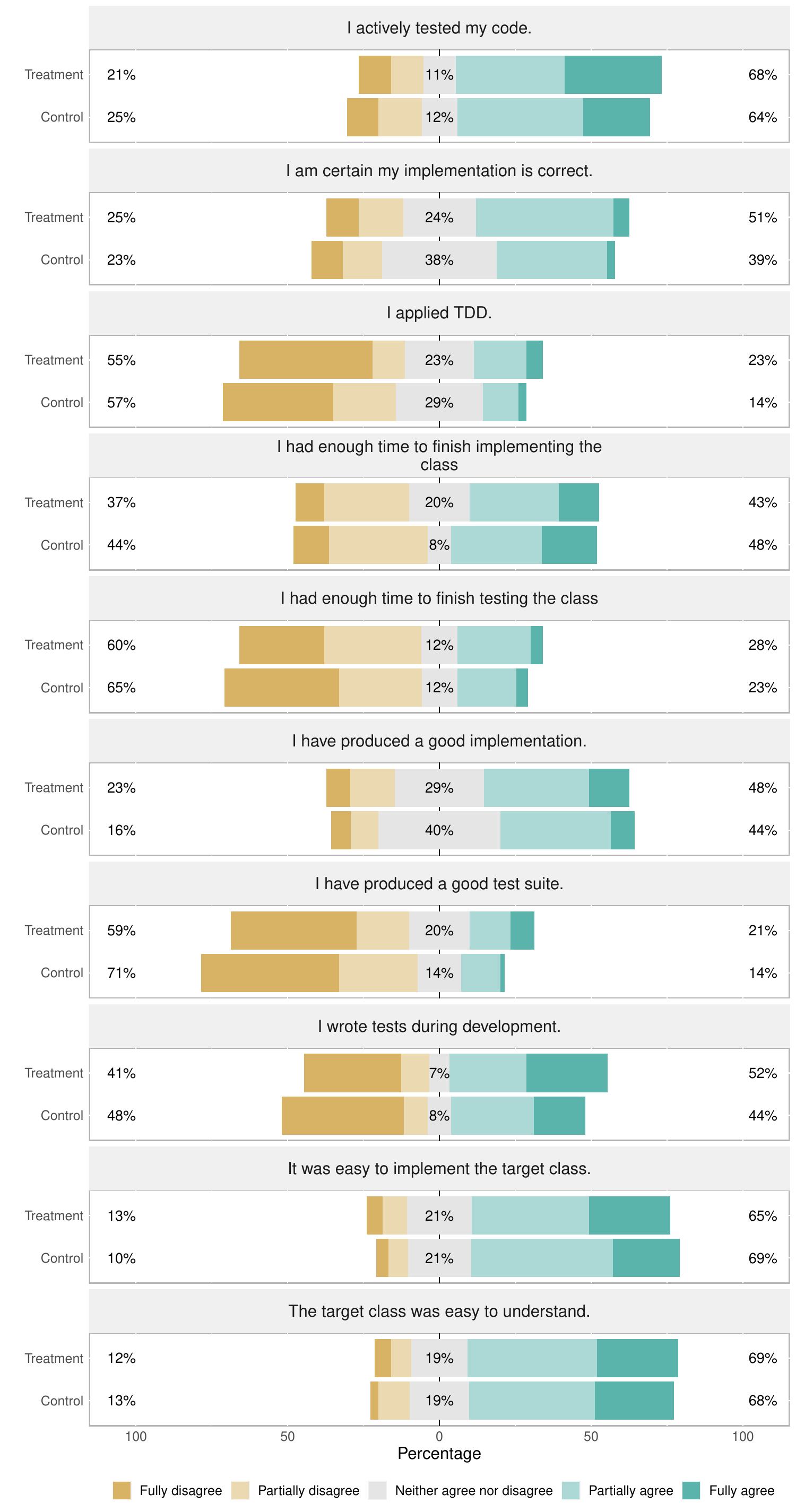}
	\caption{General survey responses -- ranging from negative on the left to positive on the right}
	\label{fig:exitall}
\end{figure}

%\todo{What is "the original set"?} 
We carefully chose a set of 23 functions from the original GitHub \emph{Date Fns} project, prioritizing simplicity in terms of complexity and lines of code. The aim was to ensure that most students could confidently build and test these functions without feeling overwhelmed by their difficulty.

Before assigning the experiment task, we conducted a pilot study with three master's students nearing graduation. This helped validate the task's scalability. While no data was collected from them, the pilot ensured proper calibration by adjusting the assignment's complexity, excluding the three most challenging functions.

The functions' bodies were removed, leaving the requirements, function names, and parameters intact. The project incorporated complete JavaScript documentation for built-in date functions, eliminating the need for online references. Two verification methods were provided: a Jest configuration with empty test files for participants to complete and run, and a \textit{main.js} file acting like the Main class in Java, mirroring the original experiment's setup~\cite{DBLP:conf/icse/Straubinger024}.

Two sessions were held: the first with 60 participants and the second with the remaining individuals. This division allowed potential data discarding from the first day if any issues arose. Participants were split into a \control group, tackling tasks without achievements, and a \treatment group, completing tasks with the plugin enabled.

Participants from the \treatment group were instructed on how the gamification plugin works but were not shown in detail the conditions for unlocking each achievement. These were shown in an overlay box displayed on the screen when hovering the mouse over each achievement of the list.

During the experiment, a custom event logger recorded achievement states after each user interaction. An automated script committed and pushed current implementations and log files to a Git repository every minute.

Following the 150-minute implementation phase, an exit questionnaire featured general questions for both groups. The treatment group received an additional page at the end of the survey inquiring about their plugin experiences, with answers on a five-point Likert scale. An optional free-text question allowed participants to share individual comments and feedback.

%\balance

\section{Results}

While the analysis of results is ongoing, we provide a preliminary overview by addressing the challenges encountered, our problem-solving approaches, and insights from the exit survey completed by students regarding their perceived user experience.

\subsection{Survey Answers}

Based on the responses to the exit survey (\cref{fig:exitall}), the chosen target class was well-received by all groups. All groups unanimously agreed that the class was easy to comprehend and implement. However, it is noteworthy that just under half of the participants in the \treatment group and fewer than 40\% in the \control group indicated they had sufficient time to complete the implementation. Consequently, it is not surprising that only approximately 25\% in both groups felt they had enough time to conduct thorough testing of their implementations.

Interestingly, around two-thirds of all participants actively tested their code. Notably, a smaller percentage of participants wrote tests during the development phase, with 52\% in the \treatment group compared to 44\% in the \control group. Roughly 50\% of participants in both groups were confident in the correctness of their implementations. Conversely, only 21\% in the \treatment group and 14\% in the \control group were certain about the quality of their test suites. No significant differences were observed between the two groups.

\begin{figure}[t]
	\includegraphics[width=\linewidth]{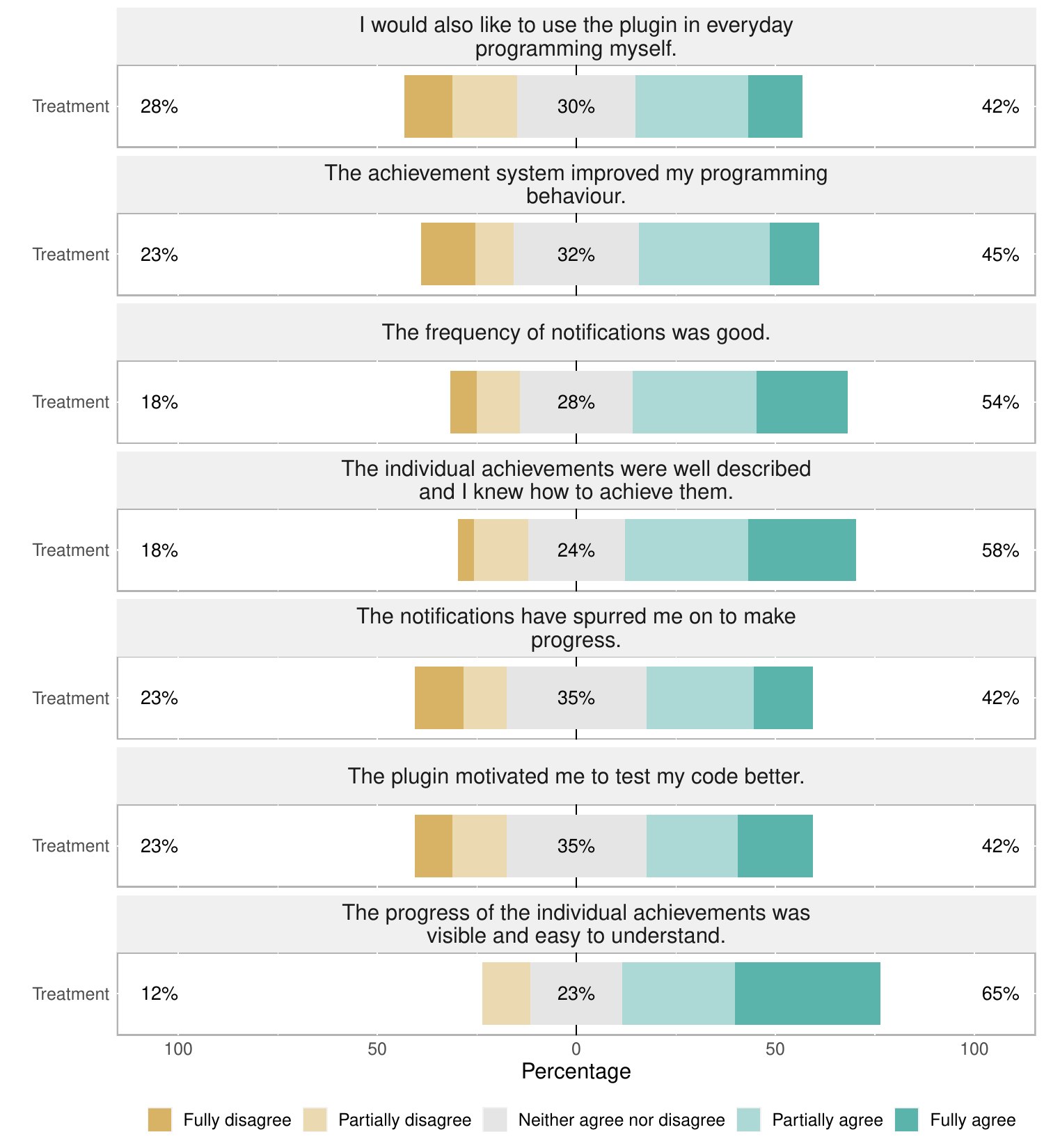}
	\caption{Answers regarding the IntelliGame plugin -- ranging from negative on the left to positive on the right}
	\label{fig:exitplugin}
\end{figure}

Concerning the responses of the \treatment group regarding \toolname, they are predominantly positive or, at worst, undecided (\cref{fig:exitplugin}). Participants demonstrated a clear understanding of the tool's descriptions and how to make progress in the presented achievements. They also appreciated the frequency of the notifications. About 40\% of the participants reported that the achievements positively influenced their testing behavior, and an equivalent percentage mentioned being motivated by both the notifications and the plugin itself. Encouragingly, 42\% of participants expressed a desire to use \toolname in their own projects.

\subsection{Problems Faced and Lessons Learned}

In the initial adaptation phase, we faced challenges while transitioning the original TypeScript project to JavaScript. Although automatic transpilation facilitated obtaining the correct JavaScript code, the hurdle lies in converting the TS configuration to a JS-compatible one. Since the original project used another testing framework, i.e., Karma, we had to modify the configuration to fit Jest.%\todo{It never really becomes clear why you need to look at TypeScript projects in the first place?}

%\todo{What file}
Another challenge arose when configuring the \textit{main.js} file for manual function assessment. Unlike Java, where a Main class defines the starting point, JavaScript, especially in the Node framework, lacks that. So, in order to reproduce the same setting as the original validation experiment~\cite{DBLP:conf/icse/Straubinger024}, we had to create a custom file for this purpose. %\todo{Why is such a starting point needed?}

Throughout the experiment, additional issues surfaced and required adaptation of the plugin. In the first session, participants could only choose between running the project via tests or main execution due to the provided configuration. Despite being shown how to switch between configurations, most participants stuck with one method for the entire session.

Upon reflection, we observed that the number of tasks overwhelmed participants, leading them to feel demotivated despite knowing that their performance would not be evaluated. While they found the functions appropriately challenging, the abundance of tasks influenced their survey responses negatively (e.g., rating their implementation/test suite as being not good enough and expressing dissatisfaction with the tool).

We attribute the superior performance of students in the pilot study to their greater experience and a less pressured environment. To address this, future experiments will involve students with comparable experience levels to the overall sample.
We hypothesize that incorporating gamification elements aimed at promoting TDD approach to the plugin may yield to even more encouraging results.

Minor issues included students incorrectly selecting a parent directory when opening the project, interrupting the script that committed their progress to the repository. We promptly identified and resolved these problems. General challenges with the committing process stemmed from variations in participants' laptop configurations, which we sometimes deemed unsolvable, opting to collect the final project state at the designated time.

%comparison with pilot

%opened the wrong folder
%configuration problem specific to users laptop unsolvable 

\section{Conclusions}

This paper presents the integration and empirical evaluation of \toolname, a gamification plugin originally designed for Java development, in the realm of JavaScript unit testing. The study aimed to validate \toolname's effectiveness in JavaScript, adapting it to the popular Jest testing framework. Despite challenges such as transitioning from IntelliJ Community Edition to Ultimate for JavaScript support and addressing differences in coverage information, we successfully ported 19 out of 26 achievements. The controlled experiment with 152 participants revealed mixed perceptions of \toolname's impact, with achievements influencing testing behaviour and participants' motivation. We will continue with an in-depth analysis of the experimental measures and the code written by the participants to gain more insights and to broaden our knowledge of \toolname.

\section*{Acknowledgement}

This study was carried out within the “EndGame - Improving End-to-End Testing of Web and Mobile Apps through Gamification” project (2022PCCMLF) – funded by European Union – Next Generation EU within the PRIN 2022 program (D.D.104 - 02/02/2022 Ministero dell’Università e della Ricerca). This manuscript reflects only the authors’ views and opinions and the Ministry cannot be considered responsible for them. This work is also supported by the DFG under grant \mbox{FR 2955/2-1}, ``QuestWare: Gamifying the Quest for Software Tests''.

\bibliographystyle{ACM-Reference-Format}
\bibliography{myBib}

\end{document}